\documentclass[aps,amsfonts,prl,twocolumn,showpacs]{revtex4-1}
\makeatletter
\AtBeginDocument{\let\LS@rot\@undefined}
\makeatother

\usepackage{amsmath,amssymb}
\usepackage{graphicx}
\usepackage{epsfig, color}
\usepackage{amsfonts}
\usepackage{bm}
\usepackage{enumerate}
\usepackage{color}
\usepackage[normalem]{ulem}
\usepackage[makeroom]{cancel}
\usepackage{multibib}
\usepackage{bbold}
\usepackage{latexsym}
\usepackage[breaklinks,colorlinks = true,linkcolor = red,urlcolor=cyan,citecolor=red,linktocpage=true]{hyperref}
\usepackage{MnSymbol}
\usepackage{comment}

\newcommand{\be}{\begin{equation}}
\newcommand{\ee}{\end{equation}}
\newcommand{\bent}{\begin{equation*}}
\newcommand{\eent}{\end{equation*}}

\newcommand{\vps}{\vphantom{*}}

\def\beq{\begin{equation}}
\def\eeq{\end{equation}}
\def\bea{\begin{eqnarray}}
\def\eea{\end{eqnarray}}

\usepackage{pdfpages} 
\makeatletter
\AtBeginDocument{\let\LS@rot\@undefined}
\makeatother

\begin{document}

\title{Diffusive hydrodynamics from integrability breaking}

\author{Aaron J. Friedman$^{1,2}$, Sarang Gopalakrishnan$^{3}$, and Romain Vasseur$^{2}$}
\affiliation{$^1$Department of Physics and Astronomy, University of California, Irvine, CA 92697, USA \\
$^2$Department of Physics, University of Massachusetts, Amherst, Massachusetts 01003, USA \\
$^3$Department of Physics and Astronomy, CUNY College of Staten Island, Staten Island, NY 10314;  Physics Program and Initiative for the Theoretical Sciences, The Graduate Center, CUNY, New York, NY 10016, USA}

\begin{abstract}

We describe the crossover from generalized  
to conventional hydrodynamics in nearly integrable systems. Integrable systems have infinitely many conserved quantities, which spread ballistically in general. When integrability is broken, only a few of these conserved quantities survive. The remaining conserved quantities are generically transported diffusively; we derive a compact and general diffusion equation for these. The diffusion constant depends on the matrix elements of the integrability-breaking perturbation; for a certain class of integrability-breaking perturbations, including long-range interactions, the diffusion constant can be expressed entirely in terms of generalized hydrodynamic data. 

\end{abstract}

\maketitle

Hydrodynamics describes how many-body systems evolve from local to global equilibrium~\cite{landau1987fluid}. It can be regarded as an effective field theory for the transport of conserved quantities or other slow modes, assuming that all other modes relax parametrically faster. Hydrodynamics applies in a variety of contexts~\cite{2018arXiv180509331G}, 
from traditional classical fluid dynamics to 
quark-gluon plasmas~\cite{2010qgp4.book..207T,2017arXiv171205815R}, black hole physics~\cite{doi:10.1146/annurev.nucl.57.090506.123120,2011arXiv1107.5780H}, and electron fluids in graphene and ${\rm PdCoO}_2$~\cite{Bandurin1055,Crossno1058,Moll1061}.  

In one dimension, many paradigmatic models of quantum many-body physics---such as the Hubbard, Heisenberg, and Lieb-Liniger models---are integrable~\cite{Calabrese:2006, sirker:2010, PhysRevLett.106.217206, PhysRevLett.110.257203, PhysRevLett.113.117202, PhysRevLett.115.157201, PhysRevLett.106.217206, PhysRevLett.110.257203, PhysRevLett.113.117202,PhysRevLett.113.117203, PhysRevLett.115.157201, 2016arXiv160300440I, 1742-5468-2016-6-064001,1742-5468-2016-6-064002,1742-5468-2016-6-064010,1742-5468-2016-6-064007,PhysRevB.89.125101}. These models approximately describe experiments in quasi-one-dimensional materials and ultracold atomic gases~\cite{kinoshita, gring, PhysRevLett.110.205301, vidmar2015, aidelsburger2018, tang2018, erne2018, zundel2018,PhysRevLett.122.090601,hess2019heat}. Thus, approximate integrability is of wide experimental relevance. In nearly integrable systems, the short-time dynamics 
are integrable, feature infinitely many conservation laws, and are described by the recently developed framework of generalized hydrodynamics (GHD)~\cite{Doyon,  Fagotti, SciPostPhys.2.2.014, PhysRevLett.119.020602,  BBH0, BBH,PhysRevLett.119.020602, GHDII, doyon2017dynamics, solitongases,PhysRevLett.119.195301,2016arXiv160408434Z, PhysRevB.96.081118,PhysRevB.97.081111, dbd1, ghkv, dbd2, horvath2019euler, PhysRevB.100.035108,2019arXiv190601654B}; at sufficiently long times, however, the dynamics are chaotic, feature 
finitely many conservation laws, and are typically described by conventional hydrodynamics. In integrable systems, transport is generically ballistic~\cite{PhysRevLett.119.020602}, although there are various limits  
that exhibit more exotic behavior~\cite{1742-5468-2009-02-P02035,lzp,sanchez2018anomalous, idmp, gv_superdiffusion,agrawal2019,gvw,PhysRevLett.123.186601,PhysRevLett.122.210602, manas2019, 2019arXiv190905263A,2019arXiv191008266B}; in conventional hydrodynamics, one expects \emph{diffusion},
unless the system possesses Galilean or Lorentz invariance~\cite{PhysRevLett.108.180601,PhysRevA.88.021603, Spohn2014, spohn_nlfhd,PhysRevA.92.043612, PhysRevE.99.012124,Das:2019aa}. The timescales governing the crossover between these two regimes have recently been explored both experimentally and numerically~\cite{tang2018,PhysRevX.9.021027}, and have been shown to match a  
Fermi Golden Rule (FGR) prediction, with matrix elements evaluated via exact numerical diagonalization on small systems. However, except in noninteracting and weakly interacting models~\cite{landau1981course,PhysRevLett.96.067202, PhysRevX.5.041043, PhysRevLett.115.180601,PhysRevB.94.245117,2017arXiv171100873C}, the nature of relaxation and the transport coefficients governing the long-time hydrodynamics have not been investigated (see, {\it e.g.}, Refs.~\citenum{PhysRevE.81.036206, PhysRevLett.120.070603, sanchez2018anomalous, PhysRevLett.123.240603,Biella:2019aa} for recent numerical studies). The existing perturbative results do not apply to many of the experimentally relevant settings, such as the Heisenberg and Fermi-Hubbard model, which are, in general, strongly interacting. Moreover, recent results suggest that anomalous transport might survive integrability breaking up to long times~\cite{PhysRevLett.120.164101, PhysRevLett.123.186601,2019arXiv190409287B,dupont_moore}, and it is crucial to construct a framework that captures how anomalous transport features due to integrability cross over to ordinary diffusion at long times. 

In this work we develop a framework for computing relaxation and diffusion in nearly integrable systems, building on GHD. A central result of this work is a compact formula for the diffusion constant in nearly integrable systems with one (or a few) residual conservation laws. The specifics of the integrability-breaking mechanism enter this formula through a set of microscopic rates that govern the decay of the approximately conserved quantities. In general, these rates depend on the microscopic mechanism of integrability breaking. However, for integrability-breaking perturbations that are spatially slowly varying (e.g., smooth potentials and long-range interactions), these rates can themselves be expressed in terms of GHD data---in these cases, the diffusion constant can be fully expressed in terms of GHD data. Having introduced these general results using both the Kubo formula and a gradient expansion of the hydrodynamic equations, we apply them to some specific systems in which the physics is particularly transparent. 

\emph{GHD Boltzmann equation}.---Integrable systems have extensively many conserved quantities and stable, ballistically propagating quasiparticles, unlike quantum chaotic systems. Within GHD~\cite{Doyon,Fagotti}, dynamics can be captured by a ``Bethe-Boltzmann'' equation for the density of quasiparticles, $\rho^{~}_\lambda(x,t)$, 
with a given quantum number (rapidity), $\lambda$:
\beq
\partial^{\vps}_t \rho_\lambda + \partial^{\vps}_x \left( v^{\rm eff}_\lambda [\rho^{\vps}]  \rho^{\vps}_\lambda \right) = \mathcal{I}^{\vps}_{\lambda} [\rho]~.~~~ \label{eqGHD1}
\eeq
where the effective group velocity $v^{\rm eff}_\lambda [\rho^{\vps}] $ of quasiparticle type $\lambda$ is a functional of the densities of all quasiparticle types. The effective  
velocity can be computed from the thermodynamic Bethe ansatz (TBA) solution for the local state of the model,~$\rho^{\vps}_\lambda(x,t)$~\cite{PhysRevLett.113.187203,Doyon,  Fagotti,2019arXiv190807320B}. 
Note that the quantum numbers $\lambda$ may be either discrete or continuous. Intuitively, Eq.~\eqref{eqGHD1} with a vanishing righthand side $\mathcal{I}^{~}_{\lambda} [\rho]=0$ is a kinetic equation that describes the ballistic propagation of the quasiparticles (solitons), which scatter elastically and delay one another through Wigner time delays~\cite{solitongases,BBH}, leading to a state-dependent velocity  $v^{\rm eff}_\lambda [\rho] $. Since scattering processes in integrable systems factorize, this kinetic equation remains valid even if the quasiparticle gas is not dilute. 

Breaking integrability endows this equation with a righthand side, $ \mathcal{I}^{~}_{\lambda} [\rho]$, which accounts for the scrambling of the quasiparticle quantum numbers (see also~Ref.~\citenum{2017arXiv171100873C}).  For simplicity, we restrict our consideration to fluctuations above homogeneous background states, $\rho (x,t)= \rho^{\star} + \delta \rho (x,t)$. Since the dynamics of interest are not strictly integrable, the steady state, $\rho^{\star}$, will, in general, describe a standard thermal Gibbs ensemble. By definition, the righthand side of the Boltzmann equation vanishes 
for $\rho = \rho^{\star}$, so one can write $\mathcal{I}^{~}_{\lambda} [\rho]= - \int d \theta  \Gamma^{~}_{\lambda, \theta}  \delta \rho^{~}_\theta (x,t)$, where $\Gamma^{~}_{\lambda,\theta} \equiv -\left. \delta \mathcal{I}^{~}_{\lambda} / \delta \rho^{~}_{\theta} \right|^{~}_{\rho = \rho^{\star}}$. The linearized version of~Eq.~\eqref{eqGHD1} then reads
\beq
\partial^{\vps}_t   \delta \rho^{\vps}_\lambda + \partial^{\vps}_x   ({\bf A}  \delta \rho)^{\vps}_\lambda =- ({\bf \Gamma}  \delta \rho)^{\vps}_\lambda~,~~~ \label{eqGHD2}
\eeq
where ${\bf A}$ and ${\bf \Gamma}$ are operators acting in rapidity space, \emph{e.g.},  as $({\bf \Gamma} \delta \rho)^{~}_\lambda = \int d \theta  \Gamma^{~}_{\lambda,\theta}  \delta \rho^{~}_\theta (x,t)$. 
The matrix ${\bf A}$ is position independent: its expression in terms of the background state, $\rho^\star$, is known exactly in GHD~\cite{GHDII} and its eigenvalues are the effective velocities $v^{\rm eff}_\lambda[\rho^\star]$; the corresponding  eigenvectors are the normal modes of GHD. 

The densities of conserved quantities are given in terms of the quasiparticles by
\beq
q^{\vps}_m(x,t) = \int d\lambda  h^{\vps}_m(\lambda)  \rho^{\vps}_\lambda(x, t)~,~~~ \label{eqCharge}
\eeq
where $h$ is the charge carried by a quasiparticle with rapidity $\lambda$. The expectation value of the conserved charge $ {\hat Q}_m$ in the generalized Gibbs ensemble (GGE)~\cite{1742-5468-2016-6-064007} corresponding to the background state, $\rho^{\star}$, is then given by $Q^{~}_m \equiv \langle {\hat Q}_m \rangle = \int dx   q^{~}_m(x)$. 
In the charge basis, the deviation of the conserved charges from their background values follows from Eq.~\eqref{eqGHD2}: $\partial^{~}_t \delta q^{~}_n + A^{~}_{nm} \partial^{~}_x \delta q^{~}_m = -\Gamma^{~}_{nm} \delta q^{~}_m$. Henceforth, repeated indices are implicitly summed over, and ${\bf A}$ and ${\bf \Gamma}$ are now 
written in the (complete) charge basis~\footnote{We are free to choose superpositions of the $Q_m$ for which the functions $h_m(\lambda)$ have a simple completeness relation.}. One has $A^{~}_{nm} = \int d \lambda  d \theta  h^{~}_n(\lambda)  A^{~}_{\lambda , \theta}  h^{~}_m(\theta) $ (and similarly for $\Gamma^{~}_{nm}$). Integrating this equation over position, $x$,  one finds for ${\bf \Gamma} \neq 0$ that the charges decay as
\beq
\delta\dot{ Q}^{\vps}_m =- \Gamma^{\vps}_{mn}  \delta Q^{\vps}_n~.~~ \label{eqQdecay}
\eeq
The eigenvalues of ${\bf \Gamma}$ give the decay rates of the quantities $\{ Q^{~}_m\}$ that are conserved when ${\bf \Gamma} = 0$; for ${\bf \Gamma} \neq 0$, the (decaying) eigenmodes of ${\bf \Gamma}$ are linear combinations of these $\{ Q^{~}_m\}$~\footnote{We require ${\bf \Gamma}$ to be positive semidefinite: its eigenvalues, $\{ \gamma \}$ obey $\mathrm{Re}(\gamma) \geq 0$.}. 
Any residual conserved quantities in the nonintegrable system, e.g., energy or particle number, correspond to zero modes of $\bf{\Gamma}$. In what follows, Greek characters denote residual conserved charges and Roman characters denote charges that decay when integrability is broken.

\emph{Kubo formula}.---We now compute the linear response d.c. conductivity tensor, $\sigma^{~}_{\alpha\beta}$, of the residual conserved charges using the Kubo formula
\beq
\sigma^{\vps}_{\alpha \beta} = \frac{1}{L} \int_0^\infty dt   \langle   {\hat J}_\alpha(t)  {\hat J}_\beta(0) \rangle, \label{eqKubo}
\eeq
evaluated in the GGE corresponding to the background state, $\rho^{\star}$, 
where $L$ is the system size and ${\hat J}^{~}_\alpha = \int dx  {\hat j}^{~}_\alpha$ is the global current associated with the conserved charge ${\hat Q}^{~}_\alpha$. 
In the integrable limit (${\bf \Gamma}=0$), one can write $J^{~}_\alpha = J_\alpha^{\mathrm{Euler}} + J_\alpha^{\mathrm{fast}}$. When ${\bf \Gamma} = 0$, the first term never decays because it can be decomposed onto conserved charges, $J_\alpha^{\mathrm{Euler}} = A^{~}_{\alpha n} Q^{~}_n$, where $A^{~}_{nm} = \partial J^{~}_n /\partial  Q^{~}_m$ 
are the components of ${\bf A}$ evaluated in the steady state, $\rho^\star$. 
The remaining fast components of the current generically relax on some characteristic timescale and give rise to diffusive and higher-order corrections to ballistic transport~\cite{dbd1,ghkv, dbd2,gv_superdiffusion}. 

On timescales at which the fast components have relaxed, 
one can take $J^{~}_\alpha \simeq A^{~}_{\alpha n}  Q^{~}_n$ in Eq.~\eqref{eqKubo}~\footnote{This Euler-scale relation is, strictly speaking, valid for the expectation values in the GGE, but is not an operator identity; hence the lack of hats.}. 
This yields $\langle  {\hat J}^{~}_\alpha(t)  {\hat J}^{~}_\beta(0) \rangle /L= A^{~}_{\alpha n}   A^{~}_{\beta m}  C^{~}_{nm} + \dots$ at long times, where the matrix elements $C^{~}_{nm} = 
\langle  \delta  {\hat Q}^{~}_n \delta {\hat Q}^{~}_m  \rangle / L$
encode the equilibrium fluctuations of the conserved charges~\footnote{Because of the way conserved charges are defined, 
the matrix ${\bf C}$ is nontrivial  even in the infinite-temperature thermal state.}, and are known exactly from TBA~\cite{PhysRevB.54.10845}. Thus, when ${\bf \Gamma}=0$, integrable dynamics generically lead to a nonzero value of the Kubo correlator and a Drude weight $D^{~}_{\alpha \beta} = \lim_{t \to \infty} \langle  {\hat J}^{~}_\alpha(t)  {\hat J}^{~}_\beta(0) \rangle /L = ({\bf A C A}^T)^{~}_{\alpha \beta}$~\cite{GHDII}. This ballistic contribution to transport follows naturally from the overlap between currents and conserved charges, which prevents the currents from decaying at long times. 

\emph{Conductivity tensor}.---When ${\bf \Gamma} \neq 0$ all but a few charges decay 
according to Eq.~\eqref{eqQdecay}, and one expects the currents to relax fully, giving rise to diffusive hydrodynamics. 
We assume that the currents are not modified by the integrability-breaking perturbation, which is justified perturbatively.
The autocorrelator in Eq.~\eqref{eqKubo} then relaxes in two stages: the fast component relaxes on a timescale of order unity and the Euler-scale component decays on a much longer timescale set by ${\bf \Gamma}$ (rather than persist indefinitely).
At long times, one can ignore the contributions from the fast part (as before), 
which is subleading in ${\bf \Gamma}$, and expand the currents in terms of the slowly relaxing charges, $\delta Q^{~}_n(t) = [\exp(-{\bf \Gamma} t)]^{~}_{nm}  \delta Q^{~}_m(0)$ to recover $\langle  {\hat J}^{~}_\alpha(t)  {\hat J}^{~}_\beta(0) \rangle /L  = A^{~}_{\alpha n}   A^{~}_{\beta m}  [{\rm e}^{-{\bf \Gamma} t}]^{~}_{n k}  C^{~}_{km} + \dots$, where neglected terms include nonhydrodynamic modes that relax at a rate much faster than ${\bf \Gamma}$. Using the Kubo formula, Eq.~\eqref{eqKubo}, 
gives the d.c.~conductivity tensor
\beq
\sigma^{\vps}_{\alpha \beta} = ({\bf A}  {\bf \Gamma}^{-1}  {\bf A })^{\vps}_{\alpha\gamma}  C^{\vps}_{\gamma \beta}~,~~ \label{eqSigma}
\eeq
where unlabeled matrix products may be evaluated in either the charge or quasiparticle basis, but are restricted to the decaying modes --- this projection onto nonconserved charges ensures that the inverse ${\bf \Gamma}^{-1}$ is well defined. Thus, 
$\sigma^{~}_{\alpha \beta}$ is nonsingular and the d.c.~limit is well-defined unless a current, $J^{~}_\alpha$, of a residual conserved 
charge, $Q^{~}_\alpha$, itself has some overlap with residual conserved charges, in which case $q^{~}_\alpha$ will  spread ballistically even when integrability is broken. 
We also used ${\bf AC} = {\bf CA}^T$~\cite{spohn1991large}, and that $C^{~}_{\alpha n} = 0$ between quantities that are residually conserved for ${\bf \Gamma} \neq 0$ and those that are not, as the latter decay to zero.
This means that the decaying and conserved charges belong to orthogonal subspaces under the hydrodynamic inner product defined by the matrix ${\bf C}$. (If they did not, one could use this nonzero overlap to prove a Mazur bound, contradicting the assumption that these charges indeed decay~\cite{VirPrivComm}.)

Eq.~\eqref{eqSigma} is a central result of this work: it expresses the conductivity tensor entirely in terms of GHD data and the rate matrix, ${\bf \Gamma}$, governing the decay of the $\{ Q^{~}_m\}$. Intuitively, this describes a generalized Drude formula: in the presence of integrability-breaking perturbations, the Drude weight for ${\bf \Gamma}=0$ is broadened into Lorentzians of width $\sim \Vert {\bf \Gamma} \Vert$ in the a.c.~conductivity. 
Importantly, ${\bf A}$ and ${\bf C}$ are known exactly for integrable systems, and we will discuss below how ${\bf \Gamma}$ can be obtained in some cases from GHD data. Note that ${\bf \Gamma}$ can be efficiently inverted numerically---such kernel operator inversions are routinely performed in the solution of TBA equations.

We remark that diffusive corrections to ballistic transport in the integrable limit (${\bf \Gamma}=0$) are negligible compared to~Eq.~\eqref{eqSigma} in the limit where the integrability-breaking perturbation is small, \emph{unless} the integrable model itself exhibits superdiffusion or diffusion if the Drude weight happens to vanish~\cite{PhysRevLett.119.020602}. We will briefly return to this case below. Finally, if the spectrum of ${\bf \Gamma}$ is gapless (\emph{i.e.}, if it has eigenvalues arbitrarily close to zero) then ballistic transport may result in \emph{anomalous} diffusion upon breaking integrability.

\emph{Diffusive hydrodynamics}.---As the matrix ${\bf C}$ is also a susceptibility matrix, one can  use a generalized Einstein relation to extract the diffusion matrix from Eq.~\eqref{eqSigma}, 
\beq
{\cal D}^{\vps}_{\alpha \beta} = ({\bf A}  {\bf \Gamma}^{-1}  {\bf A})^{\vps}_{\alpha \beta}~,~~~ \label{eqDiffusion}
\eeq
which will depend on the Lagrange multipliers, $\{ \beta^{~}_\gamma \}$, of the GGE corresponding to the charges preserved by the integrability-breaking perturbation. Although Eq.~\eqref{eqDiffusion} derives from linear response, ${\cal D}^{~}_{\alpha \beta}$ can be used as a transport coefficient to formulate a fully nonlinear hydrodynamic equation describing the dynamics at late times $t \gg \tau \sim ||{\bf \Gamma}^{-1}||$,
\beq
\partial^{\vps}_t  \delta q^{\vps}_\alpha = \partial^{\vps}_x  \left( {\cal D}^{\vps}_{\alpha \beta}  [\lbrace q^{\vps}_\gamma \rbrace ]   \partial^{\vps}_x  \delta q^{\vps}_\beta \right)~,~~ \label{eqDiffusioneq}
\eeq
where the Lagrange multipliers, $\{ \beta^{~}_\gamma \}$, have been replaced by the expectation values of the conserved charges. (In principle Eq.~\eqref{eqDiffusioneq} also includes a noise term, not shown, whose strength is fixed by the fluctuation-dissipation theorem.)

A more direct way to derive this diffusion equation in the linear response regime is as follows. For concreteness, we consider the case with a single residual conserved charge, $q^{~}_0$. The Euler-scale hydrodynamic equations are
\bea\label{eq9}
\partial^{\vps}_t  \delta q^{\vps}_0 + A^{\vps}_{0n}  \partial^{\vps}_x \delta q^{\vps}_n & = & 0~, \\
\partial^{\vps}_t  \delta q^{\vps}_{n} + A^{\vps}_{nm}  \partial^{\vps}_x  \delta q^{\vps}_m & = & - \Gamma^{\vps}_{nm}  \delta q^{\vps}_m~, \quad n \neq 0~.~~ \nonumber
\eea
To leading order in the gradient expansion, we may drop derivatives of  the $\delta q^{~}_n, \forall n \neq 0$ in the second equation, yielding $\delta q^{~}_n = -\Gamma^{-1}_{nm}   A^{~}_{m 0}   \partial^{~}_x  \delta q^{~}_0 + \dots$ Substituting this into the first equation recovers the diffusion constant ${\cal D}^{~}_{00} = A^{~}_{0n}   \Gamma^{-1}_{nm}   A^{~}_{m 0}$, consistent with the Kubo result, Eq.~\eqref{eqDiffusion}~\cite{SuppMat}.
We emphasize that here diffusion arises from ``integrating out'' slow but nonconserved degrees of freedom, and is dramatically different from the diffusive corrections that arise in integrable systems (${\bf \Gamma}=0$) due to the fluctuations of ballistically propagating modes~\cite{ghkv,2019arXiv191101995M,2019arXiv191201551D}. 

Note that the preceding arguments do not rely on spatial locality of ${\bf \Gamma}$, and, in fact, generalize to the case wherein ${\bf \Gamma}$ is a spatially nonlocal kernel. In that case, the diffusion equation takes the form $\partial^{~}_t  q^{~}_0(x,t) = \partial^{~}_x  \left\{ \int dy  {\cal D}(x - y)  \partial^{~}_y  q^{~}_0(y) \right\}$, where ${\cal D}(x-y) \equiv A^{~}_{0n}(x)  \Gamma^{-1}_{nm}(x - y)  A^{~}_{m0}(y)$. When ${\cal D}(x-y)$ is sufficiently long ranged, the nature of the hydrodynamics might change, though we will not consider this case in detail.

\emph{Hydrodynamic projections and general operators}.---In the discussion above, we analyzed the dynamics of current autocorrelators. However, the essential ingredient---namely, the separation of an operator into fast and slow components, where the latter correspond to overlaps of the operator with almost conserved charges---is true for \emph{any} operator. Thus, the analysis above directly generalizes to the autocorrelation function of an arbitrary ``global'' operator $\hat{\mathcal{O}} = \sum_i \hat{\mathcal{O}}^{~}_i$, via the formalism of hydrodynamic projections (see, \emph{e.g.},~Ref.~\citenum{10.21468/SciPostPhys.5.5.054}). 
For simplicity, we assume that $\langle  \hat{\mathcal{O}}  \rangle = 0$ in the GGE associated with the background state, $\rho^{\star}$. The projection of $\hat{\mathcal{O}}$ onto 
a slow (but nonconserved) charge, $Q^{~}_m$ can be expressed as
\beq
 \langle  \mathcal{O}   |  Q^{\vps}_m \rangle = - \partial^{\vps}_{\beta_m}  \langle  \hat{\mathcal{O}}   \rangle\big\vert^{\vps}_{\beta^{\vps}_m = 0}~,~~
\eeq
When ${\bf \Gamma} = 0$, one can use the TBA formalism to compute expectation values for any value of the chemical potential, $\beta^{~}_m$, associated with the charge $Q^{~}_m$, and can thus evaluate the projection for 
sufficiently simple operators. 
It readily follows that 
\beq
\langle  \hat{\mathcal{O}} (t)  \hat{\mathcal{O}} (0)  \rangle =  \langle  {\mathcal{O}}  |  Q^{\vps}_m   \rangle  C^{-1}_{mn}  [e^{-{\bf \Gamma} t}]^{\vps}_{np}   \langle Q^{\vps}_p  |  \mathcal{O}   \rangle~.~~~
\eeq
For current operators $J^{~}_\alpha$, the hydrodynamic projection $ \langle  J^{~}_\alpha  |  Q^{~}_n  \rangle = B^{~}_{\alpha n} = - \partial J^{~}_\alpha / \partial \beta^{~}_n$ defines the matrix ${\bf B}={\bf A  C}$~(by the chain rule)~\cite{spohn1991large}, which recovers Eq.~\eqref{eqSigma}.

\emph{Transition rates}.---So far, we have expressed the behavior of autocorrelation functions in terms of GHD data and the matrix ${\bf \Gamma}$, which describes the decay of the conserved charges due to collisions. We now discuss how one can compute ${\bf \Gamma}$ perturbatively. From Fermi's Golden Rule, the RHS of Eq.~\eqref{eqGHD1} takes the general schematic form
\bea\label{fgr}
\mathcal{I}^{\vps}_\lambda & = &   \int\! \prod_{ij} d \alpha^{~}_i  d\beta^{~}_j  \Big( \prod_{ij} \rho^{~}_{\alpha^{~}_i}  \rho^h_{\beta^{~}_j}  \rho^h_{\lambda}  \left| M^{~}_{\{\alpha^{~}_i \} \rightarrow \{ \beta^{~}_j, \lambda \}}\right|^2 \nonumber \\ && \qquad\qquad - \rho^{~}_\lambda  \prod_{ij} \rho^{~}_{\beta^{~}_j} \rho^h_{\alpha^{~}_i}  \left| M^{~}_{\{\lambda, \beta^{~}_j \} \rightarrow \{\alpha^{~}_i \}}\right|^2 \! \Big)~,~
\eea
where $ \rho^h_{\alpha}$ is the density of holes with rapidity $\alpha$. We also introduce the density of states, $\rho^{\rm tot}_\lambda = \rho^{~}_\lambda + \rho_\lambda^h$, and the occupation factor, 
$n^{~}_\lambda = \rho^{~}_\lambda/ \rho^{\rm tot}_\lambda$ . Here, $M$ denotes matrix elements of the integrability-breaking perturbation between eigenstates of the integrable system. The first term corresponds to scattering particles into the quasiparticle state $\lambda$, and the second to scattering them out. The scattering can happen in various permutations, which must be summed over. In general, the matrix elements that enter this expression must be derived from microscopics; however, in some cases, they can be expressed in terms of GHD data in the hydrodynamic limit. 

As a simple example we consider an interacting one-dimensional Bose gas (the Lieb-Liniger model with particle mass $m$) subject to a weak, smoothly varying time-dependent potential coupled to one of the charges~\cite{2019arXiv190601654B}, i.e. $V(x)   \eta(t)  {\hat \rho}(x)$, with ${\hat \rho} = {\hat q}^{~}_0$, the quasiparticle density. (For the less trivial case of an interaction with a smooth kernel see~\cite{SuppMat}.) A key observation~\cite{panfil2018,dbd2} is that, at long wavelengths, the dominant matrix elements of $V(x)  {\hat \rho}(x)$ are those that rearrange the fewest quasiparticles, regardless of interaction strength. Thus we can restrict to one-particle-hole excitations, for which the matrix elements are given by $\langle  m  |  \hat q^{~}_0  |  m; \{ \lambda \rightarrow \theta \}  \rangle = h_0^{\mathrm{dr}}(\lambda)$~\cite{dbd2}, with $h_0^{\mathrm{dr}}$ the ``dressed'' charge, $h_0^{\mathrm{dr}} = ({\bf 1} +{\bf n}   {\bf K})^{-1}  h^{~}_0$ where $h^{~}_0(\lambda)=1$ for particle number, ${\bf K}$ is the scattering kernel of the model, and ${\bf n}$ acts diagonally in rapidity space as the occupation factor, $n^{~}_{\lambda}$. 
We find that
\bea
\mathcal{I}^{\vps}_\lambda & = &  \rho^{\rm tot}_\lambda   \int d\varphi  \left|  \tilde V(k^{~}_{\lambda+\varphi} -k^{~}_{\lambda} )\right|^2  \left|\tilde{\eta} (\varepsilon^{~}_{\lambda + \varphi} - \varepsilon^{~}_\lambda)\right|^2   \\ & &   \rho^{\rm tot}_{\lambda+\varphi}  h_0^{\mathrm{dr}}(\lambda)  h_0^{\mathrm{dr}}(\lambda + \varphi)  \big[ n^{~}_{\lambda +\varphi} (1-n^{~}_\lambda) - n^{~}_\lambda (1 -n^{~}_{\lambda +\varphi} ) \big]~,~ \nonumber
\eea
where $\tilde V$, $\tilde \eta$ denote Fourier transforms and $\varepsilon^{~}_\lambda$ and $k^{~}_\lambda$ are, respectively, the dressed energy and momentum of the excitation, satisfying  $\varepsilon' = (E')^{\rm dr}$ and $k' = (P')^{\rm dr}$, with $E(\lambda) = m  \lambda^2 / 2$ and $P(\lambda) = m  \lambda$ the single-particle energy and momentum (entering Eq.~\eqref{eqCharge}). 
If we fix the background state, this is similar to the scattering of free fermions with charge $h^{\mathrm{dr}}_0$. 
This dependence on the dressed charge is also seen in the more complicated case of slowly varying interactions~\cite{SuppMat}. 
However, our assumptions fail for many important types of scatterings within an integrable system, such as the decay of one quasiparticle type into another or Umklapp scattering of quasiparticles; incorporating these is an interesting topic for future work.

\emph{Examples}.---We now comment on the physical significance of our results by considering  
several specific cases. First, consider an interacting Bose gas in one dimension with particle mass $m$. By Galilean invariance the current corresponding to the boson density, $q^{~}_0$, is the momentum, $j^{~}_0 = q^{~}_1$. Suppose that the momentum distribution relaxes to a Gaussian on a timescale $\tau$ (\emph{e.g.}, due to point scatterers). Then Eq.~\eqref{eqSigma} predicts a conductivity $\sigma = \tau  \chi  A^{~}_{01}  A^{~}_{10}$ for the boson density. We have $A^{~}_{01}=1$ by Galilean invariance, and $A^{~}_{10}= \chi^{-1}  n/m$ (see {\it e.g.}~\cite{VKM}). We thus recover the Drude formula $\sigma = \tau  n/m$ (momentum relaxation time times Drude weight). A similar result would apply to energy transport in the spin-$1/2$ XXZ chain with pointlike scatterers. 

Next, we turn to cases in which corrections to the Euler scale are large and potentially even divergent. An example is the spin-$1/2$ XXZ chain with easy-plane anisotropy. For concreteness, we consider subjecting this system to slowly varying noise, as 
before. In this model, the Drude weight varies discontinuously with the anisotropy parameter~\cite{PhysRevLett.82.1764, karraschdrude, Prosen20141177,PhysRevB.96.081118,PhysRevLett.111.057203,PhysRevLett.119.020602, urichuk2019spin}, and, concomitantly, the low-frequency response is anomalous, so that in the integrable limit, 
for generic anisotropy, one has $\sigma(\omega) = D \delta(\omega) + c  \omega^{-1/2}$~\cite{2019arXiv190905263A}. Only the largest quasiparticles (``strings'') in this model are charged under the magnetization, so only these strings couple to the integrability-breaking perturbation. In the integrable limit, these strings undergo a L\'evy flight; when integrability is broken, the L\'evy flight is cut off and crosses over to diffusion with a mean free time $\tau$. The d.c. conductivity then goes as $\sigma =D  \tau + c  \sqrt{\tau}$, \emph{i.e.}, it corresponds to convolving the integrable result with a Lorentzian of width $\tau^{-1}$. Note that this result is nonanalytic in $\tau$: anomalous transport in the integrable limit can result in signatures in the nonanalytic dependence of the diffusion constant on the integrability-breaking parameter. One can try to extend this analysis to the  easy-axis regime of the XXZ model where spin transport is diffusive in the integrable limit; in this regime, none of the quasiparticles carry any dressed magnetization, thus at the present level of analysis their relaxation rates vanish. However, spin transport remains diffusive upon breaking integrability:  the spin Drude weight is zero since the quasiparticles are neutral, but the factors of dressed magnetization cancel out in eq.~\eqref{eqDiffusion}, so we predict a finite diffusion constant which can be computed by adding a small magnetic field $h$ (which makes the Drude weight and relaxation rates non-zero) and taking $h \to 0$ in~\eqref{eqDiffusion}.

\emph{Conclusion}.---This work has shown how the crossover from generalized to conventional hydrodynamics can be captured within the framework of GHD, by introducing a collision integral into the Bethe-Boltzmann equation. GHD allows one to write compact formulas for the diffusion constants of the residual conserved quantities, as well as for more general autocorrelation functions; it also gives access to the full, potentially nonlinear and spatially nonlocal, diffusion equations for the residual conserved quantities. These formulas involve hydrodynamic data as well as a matrix of quasiparticle decay rates, which (in the most general case) lies beyond the scope of GHD. Nevertheless, in certain limits where the collisions involve small momentum transfer, the rates can themselves be expressed in terms of GHD data, thus allowing for a fully GHD description of nearly integrable systems. Applying this technology to extract specific quantitative predictions for experiments, by incorporating the collision integral into the flea-gas algorithm~\cite{solitongases} for integrable dynamics, is a natural avenue for future work~\footnote{A. J. Friedman et al., in preparation.}.

\begin{acknowledgments}

\emph{Acknowledgments}.---The authors thank Philipp Dumitrescu, Vadim Oganesyan, and especially Vir Bulchandani for useful discussions. This work was supported by the National Science Foundation under NSF Grant No. DMR-1653271 (S.G.),  the US Department of Energy, Office of Science, Basic Energy Sciences, under Early Career Award No. DE-SC0019168 (R.V.),  and the Alfred P. Sloan Foundation through a Sloan Research Fellowship (R.V.). 

\end{acknowledgments}

\bibliography{refs}

%

%

\bigskip

\onecolumngrid
\newpage



\section*{Supplementary material}

# Supplemental Material for "Diffusive hydrodynamics from integrability breaking"

Aaron J. Friedman[1,2], Sarang Gopalakrishnan[3], and Romain Vasseur[2]

[1]*Department of Physics and Astronomy, University of California, Irvine, CA 92697, USA*
[2]*Department of Physics, University of Massachusetts, Amherst, Massachusetts 01003, USA*
[3]*Department of Physics and Astronomy, CUNY College of Staten Island,*
*Staten Island, NY 10314; Physics Program and Initiative for the Theoretical Sciences,*
*The Graduate Center, CUNY, New York, NY 10016, USA*

## I. SOLUTION OF THE LINEARIZED HYDRODYNAMIC EQUATIONS

Our starting point are the hydrodynamic equations:

$$\partial_t \delta q_0 + A_{0n} \partial_x \delta q_n = 0,$$
$$\partial_t \delta q_n + A_{nm} \partial_x \delta q_m = -\Gamma_{np} \delta q_p, \quad n \neq 0. \tag{1}$$

Here, $Q_0$ is conserved by the integrability breaking perturbation, while all the other charges decay slowly at a rate set by $\boldsymbol{\Gamma}$. To solve the dynamics, we invert the second equation to get

$$\delta q_n = -(\boldsymbol{\Gamma}^{-1} \partial_t + \boldsymbol{\Gamma}^{-1} \mathbf{A} \partial_x)_{n0} \delta q_0 - \sum_{m \neq 0} (\boldsymbol{\Gamma}^{-1} \partial_t + \boldsymbol{\Gamma}^{-1} \mathbf{A} \partial_x)_{nm} \delta q_m, \tag{2}$$

where all matrix products and inverses are within the non-conserved subspace $n \neq 0$. We can formally solve this equation by iterations, in the limit $\omega \tau \ll 1$ and $k\ell \ll 1$, with $\omega \sim \partial_t$ the frequency, $k \sim \partial_x$ the momentum, $\tau \sim ||\boldsymbol{\Gamma}^{-1}||$ the typical relaxation time, and $\ell = v^{\rm eff} \tau$ the mean free path with $v^{\rm eff} \sim ||\mathbf{A}||$ a velocity scale. This allows us to find a formal expression of the non-conserved charges in terms of $\delta q_0$

$$\delta q_n = \sum_{m \geq 1} (-1)^m \left[ (\boldsymbol{\Gamma}^{-1} \partial_t + \boldsymbol{\Gamma}^{-1} \mathbf{A} \partial_x)^m \right]_{n0} \delta q_0 \approx -(\boldsymbol{\Gamma}^{-1} \mathbf{A})_{n0} \partial_x \delta q_0 + \dots \tag{3}$$

to leading order in the gradient expansion. Using this expression in the continuity equation for $\delta q_0$ (1), we find

$$\partial_t \delta q_0 = (\mathbf{A} \boldsymbol{\Gamma}^{-1} \mathbf{A})_{00} \partial_x^2 \delta q_0. \tag{4}$$

As expected, we find a diffusion equation for the charge $\delta q_0$, with a diffusion constant $\mathcal{D} = (\mathbf{A} \boldsymbol{\Gamma}^{-1} \mathbf{A})_{00}$.

## II. GOLDEN-RULE RATES FOR SLOWLY VARYING INTERACTIONS

We consider integrability-breaking interactions of the form $V(x - y)\hat{q}(x)\hat{q}(y)$. We assume that the interaction is asymptotically sufficiently short-range, i.e., $V(x - y) \lesssim 1/|x - y|^2$ as $|x - y| \to \infty$; however, we also take it to be long-range enough that the associated momentum transfer is small. We also assume that most of the particles that interact are farther than a thermal correlation length away from one another, so that, between two many-body eigenstates, $\langle m|\hat{q}(x)\hat{q}(y)|n\rangle \sim \langle m|\hat{q}(x)|n\rangle \otimes \langle m'|\hat{q}(y)|n'\rangle$. In the latter expression, the quantum states are representative local eigenstates drawn from the appropriate Gibbs ensemble.

As we noted in the main text, in this long-wavelength limit, the dominant many-body matrix elements are those that involve rearranging as few particles as possible. (These are dominant in the sense that the one- and two-particle processes essentially exhaust the relevant sum rule [1].) In one dimension, elastic two-body collisions do not relax momentum or energy; thus the lowest-order usable processes are Umklapp scattering (which occurs only in lattice models, and requires large momentum transfer, thus falling outside the scope of our discussion) and diffractive three-body scattering (see [2]). We consider the latter process in the limit of small momentum transfer.

One can make up a three-body process by using either $\hat{q}$ to rearrange two quasiparticles and the other to rearrange a single quasiparticle. The matrix element of the density operator for rearranging a single quasiparticle of rapidity $\lambda$ is $h^{\rm dr}(\lambda)$, as noted in the main text. For the two-body process, the form of the matrix element in general is conjectured

to take the general form [Ref. [3], Eq. (3.49), see also [4]]

$$\langle m|\hat{q}|m;\{\theta_1^h,\theta_2^h\}\rightarrow\{\theta_1^p,\theta_2^p\}\rangle \;=\; 2\pi(\theta_1^p+\theta_2^p-\theta_1^h-\theta_2^h)\left(\frac{K^{\mathrm{dr}}_{\theta_2^h,\theta_1^h}h^{\mathrm{dr}}(\theta_2^h)}{k'_{\theta_1^h}k'_{\theta_2^h}(\theta_1^p-\theta_1^h)}+\frac{K^{\mathrm{dr}}_{\theta_1^h,\theta_2^h}h^{\mathrm{dr}}(\theta_1^h)}{k'_{\theta_1^h}k'_{\theta_2^h}(\theta_2^p-\theta_2^h)}\right.$$
$$\left.+\frac{K^{\mathrm{dr}}_{\theta_2^h,\theta_1^h}h^{\mathrm{dr}}(\theta_2^h)}{k'_{\theta_1^h}k'_{\theta_2^h}(\theta_2^p-\theta_1^h)}+\frac{K^{\mathrm{dr}}_{\theta_1^h,\theta_2^h}h^{\mathrm{dr}}(\theta_1^h)}{k'_{\theta_1^h}k'_{\theta_2^h}(\theta_1^p-\theta_2^h)}\right),\tag{5}$$

with $\mathbf{K}^{\mathrm{dr}}=(\mathbf{1}+\mathbf{nK})^{-1}\mathbf{K}$ the dressed scattering kernel. Here, $\theta_{1,2}^h$ are arbitrary but close to $\theta_{1,2}^p$, i.e., this form is reliable close to the poles; for slowly varying interactions, however, the momentum transfer is small and the matrix element is large only in the vicinity of the poles, so Eq. (5) is a valid approximation. In terms of these matrix elements, the collision integral can be expressed as

$$\mathcal{I}_\lambda \;=\; \int d\theta_0^h d\theta_1^h d\theta_2^h d\theta_1^p d\theta_2^p[\tilde{V}(k_{\theta_0^h}-k_\lambda)]^2\rho_{\theta_0^h}\rho_{\theta_1^h}\rho_{\theta_2^h}\rho_{\theta_1^p}^h\rho_{\theta_2^p}^h\rho_\lambda^h\tag{6}$$
$$\times(2\pi)^2\delta[k_{\theta_0^h}+k_{\theta_1^h}+k_{\theta_2^h}-k_{\theta_1^p}-k_{\theta_2^p}-k_\lambda]\delta[\varepsilon_{\theta_0^h}+\varepsilon_{\theta_1^h}+\varepsilon_{\theta_2^h}-\varepsilon_{\theta_1^p}-\varepsilon_{\theta_2^p}-\varepsilon_\lambda)]$$
$$\times[h^{\mathrm{dr}}(\theta_0^h)]^2\left|\langle m|\hat{q}|m;\{\theta_1^h,\theta_2^h\}\rightarrow\{\theta_1^p,\theta_2^p\}\rangle\right|^2+\text{permutations}.$$

Here, we have only shown the first in-scattering term, in which a quasiparticle is scattered into the state $\lambda$ by a three-body collision; its permutations include two other in-scattering terms, in which one of the other final states is $\lambda$, as well as three out-scattering terms in which one of the $\theta^h$ is replaced by $\lambda$. Eqs. (6) and (5) together specify the collision integral purely in terms of GHD data, albeit in a rather unwieldy form. A detailed analysis of these equations and their asymptotics is deferred to future work. A few general properties follow from the nature of these equations, however:

1. Assuming small momentum transfer across the vertex, one can Taylor expand the arguments in the $\delta$-functions so that the momentum-conserving $\delta$-function has the argument $\delta[k'_{\theta_0^h}(\theta_0^h-\lambda)+k'_{\theta_1^h}(\theta_1^h-\theta_1^p)+k'_{\theta_2^h}(\theta_2^h-\theta_2^p)]+(1\leftrightarrow 2)$, and similarly for the energy-conserving $\delta$-function. Given these two conditions, the integral (6) runs over a three-dimensional submanifold of the five-dimensional space. This three-dimensional submanifold in turn contains a two-dimensional space corresponding to forward-scattering processes; the matrix element (5) diverges in this subspace but these processes do not relax energy or momentum, and therefore must be excluded from the collision integral by some regularization procedure.

2. The matrix element depends on two dressed charges but not the third. Thus, even if a quasiparticle is neutral under $\hat{q}$, it can relax through this interaction, by scattering off a pair of charged quasiparticles.

3. The decay rates are proportional to the dressed kernel, $(K^{\mathrm{dr}})^2$, and thus to the strength of interactions in the integrable system.

---



\end{document}